\documentstyle[11pt]{article}
\newcommand {\be}{\begin{equation}} 
\newcommand{\fe}{\end{equation}}
\newcommand{\eqn}{\label}

\begin{document}

\font\sm=cmr9
\def\spose#1{\hbox to 0pt{#1\hss}}
\def\lta{\mathrel{\spose{\lower 3pt\hbox{$\mathchar"218$}}
     \raise 2.0pt\hbox{$\mathchar"13C$}}}
\def\gta{\mathrel{\spose{\lower 3pt\hbox{$\mathchar"218$}}
    \raise 2.0pt\hbox{$\mathchar"13E$}}}

\title{RENORMALIZATION Of GRAVITATIONAL 
\\ SELF INTERACTION FOR WIGGLY STRINGS}

\author{ {\bf Brandon Carter}
\\ D.A.R.C., Observatoire de Paris, 92 Meudon, France.}

\date{May, 1998; extended Sept., 1998.}

\maketitle

\noindent{\bf Abstract:}
{\it It is shown that for any elastic string model with energy density 
$U$ and tension $T$, the divergent contribution from gravitational
self interaction can be allowed for by an action renormalisation 
proportional to $(U-T)^2$. This formula is applied to the important
special case of a bare model of the transonic type (characterised by a
constant value of the product $UT$) that represents the macroscopically
averaged effect of shortwavelength wiggles on an underlying microscopic model
of the Nambu-Goto type (characterised by $U=T$).}

\hskip 1 cm
\section{Introduction}

Although not so important for lightweight cosmic strings, such as may may have
been formed later on (for example at the time of electroweak symmetry
breaking) gravitational self interaction is generally supposed to have played
an essential role in the evolution of the cosmic strings that may have been
formed earlier, at the epoch of G.U.T. symmetry breaking. In the kind of
scenario~\cite{VS94} originally proposed by Kibble an initial period during
which the main damping mechanism was the friction exerted by the ambient
thermal gas would be followed by a period during which the main damping
mechanism would have been gravitational, at least for the local kind of
strings to be considered here (in contrast with  global strings for which
axion radiation damping would have been more important). 

Nearly all the work that has been done~\cite{VS94}  on gravitational self
interaction in cosmic strings has been based on the use of a string model of
the simplest kind, namely that of Nambu-Goto, in which (with the speed of
light set to one) the energy density $U$ and the tension $T$ are both equal
to a constant $m^2$, where $m$ is a mass scale that will typically be of the
order of the mass associated with the Higgs field responsible for the
relevant vacuum symmetry breaking. The use of such a simple model will be
justifiable while gravitational self interaction is important, even in
typical cases where the strings are of the current carrying kind whose
likely relevance was first pointed out by Witten~\cite{W85}, since, at least
at the outset, the currents would be expected to be very weak so that
their effects would be relatively negligible.  

At a later stage the currents in small loops would intensify as the loops
contracted due to radiative energy loss, so that it might become necessary
to use a description based on an elastic string model of an appropriate
kind~\cite{CP95}, in terms of which it is possible to describe
oscillations~\cite{CM93,CPG97} about stationary ``vorton''
states~\cite{DS88} (whose existence would not even be possible if the
Nambu-Goto description remained valid). However by this later stage (except
in the case of an electromagnetically neutral current) the main self
interaction mechanism would no longer be gravitational but electromagnetic.
Thus so long as one has to do with a regime in which gravitation is still
the main self interaction mechanism, a Nambu-Goto description will usually
be adequate.

Although adequate in principle, an absolutely precise Nambu-Goto description
will seldom be feasible in numerical simulations, due to the enormous amount
of information that would be required for following the detailed evolution
of small scale wiggles. Some kind of approximation will therefore be needed
in pratice. Following the remark by Shellard and Allen~\cite{SA90} that in
the presence of small scale wiggles the macroscopically averaged value of
the energy density $U$ would exceed its Nambu-Goto value $m^2$, I
observed~\cite{C90} that the tension $T$ would be correspondingly
diminished, and proposed as an approximation the use of a model in which the
product of energy density and tension remained constant, 
\be UT=m^4 \, .\fe 
This relation was confirmed for a special subclass of wiggle modes by
Vilenkin~\cite{V90} and more recently for unrestricted wiggle modes by
Martin~\cite{M95}. A particularly attractive feature of this model is the
property of being transonic, in the sense that the propagation speed
\be c_{_{\rm E}}=\sqrt{T/U} \fe
of its extrinsic (tranverse) perturbations  is the same as the speed 
\be c_{_{\rm L}}=\sqrt{-dT/dU} \fe
of its longitudinal (sound type) modes. This makes it possible to
prove~\cite{C95} that such a model can match the long term evolution of the
string with any desired acuracy (depending on the resolution) and without any
accumulative error buildup, at least in flat spacetime where the dynamical
equations are exactly integrable.

For the practical description of  cosmic strings in circumstances where the
main kind of self interaction is gravitational it is therefore this transonic
model that will commonly be most appropriate. The purpose of this article is to
consider the way this model will need to be modified to allow for the dominant
effect of the self interaction, which will be divergent.

As in the simpler case of electromagnetic self interaction, gravitational self
interaction in a four dimensional spacetime background will give rise in point
particle models to pole type singularities, and in string models to logarithmic
singularities, which need to be regularised by the use of an ``ultraviolet''
cut off.  In order to do this in a systematic manner I have developed a simple
technical formalism~\cite{C97} whose application to the electromagnetic self
interaction gives results that agree with the conclusions of earlier
approaches, while providing a more convenient means of dealing with
applications to particular problems such as oscillations of a conducting string
loop about a stationary vorton state~\cite{GPB98}. The more recent application
of this formalism to the linearised gravitational case~\cite{BC95} has however
given results~\cite{C98} that deviate from what had previously been
obtained~\cite{VS94} in the only case that had been considered previously, 
namely that of a Nambu-Goto string, for which the divergent part simply 
vanishes. A detailed examination of this particular case~\cite{CB98} has 
shown that previous assertions to the contrary~\cite{VS94} were effectively 
based on the neglect of terms that would indeed be relatively small when due 
to high frequency gravitational radiation from a distant source, but that in 
the case of local self gravitation are of the same order as the other terms, 
which they finally cancel.

What will be done here  is to apply the general formula\cite{C98}
for the divergent part of the self gravitational interaction of a string to
the non trivial case of a generic elastic string model, and in particular to
the case of the transonic string model, for which it does not vanish. The non
vanishing result for the transonic string model is entirely consistent with
its interpretation as a smoothed approximation to a wiggly Nambu-Goto model
for which the corresponding short range self gravitational contribution
vanishes. The non-vanishing short range contribution in the approximate
description  represents the finite intermediate range contribution in the
underlying Nambu-Goto model. Thus the treatment provided here takes care of
everything except the very long range part of the self interaction, which will
be dynamically negligible in the short run, though it is of course very
important in the long run since it is the part that is ultimately responsible
for the radiative damping. 

Unlike the dissipative long range part, the dominant contribution considered
here is strictly conservative: one of the main results of this work is the
demonstration that for a generic (not just a transonic) elastic string model
the divergent self gravitational stress energy contribution given (``on
shell'') by the new formula\cite{C98} is derivable from a corresponding (``off
shell'') action contribution that is precisely the same as what is obtained as
the four dimensional specialisation of a more general (higher dimensional)
action formula that has recently been derived using an entirely different
approach by Buonanno and Damour~\cite{BD98}. The net effect is thus
describable as an action renormalisation, whose effect is trivial in the case
a Nambu-Goto model for which (consistently with what was suggested by
exact analytic considerations in static configurations~\cite{G98}) 
it simply vanishes, but non-trivial in the more general
string models considered here.

\bigskip
\section{Gravitational force density}

Since the cosmic strings to which this work applies will be characterised by
a gravitational coupling constant that is very small,
$\hbox{\sm G}m^2\lta 10^{-6} $ (where $\hbox{\sm G}$ is Newtons constant)
even for the heavyweight case of GUT strings, it is sufficient to base the
analysis on linearised gravitational theory, as expressed in terms of a
small but perhaps rapidly varying perturbation $h_{\mu\nu}=\delta
g_{\mu\nu}$ of a slowly varying cosmological background metric $g_{\mu\nu}$.
Moreover for the purpose of analysing the short range self interaction it
will sufficient to neglect the background curvature altogether, i.e. to take
unperturbed 4-dimensional spacetime metric $g_{\mu\nu}$ to be flat, so that
subject to the usual~\cite{G98} DeDonder gauge condition, 
$\nabla^\mu h_{\mu\nu}={1\over 2}\nabla_\nu h^\mu_{\,\mu}$,
the linearised Einstein equations for the Eulerian metric perturbation will
reduce to the well known form
\be \nabla_{\!\sigma}\nabla^\sigma h^{\mu\nu}=-8\pi 
\hbox{\sm G} \big(2\hat T^{\mu\nu}
-\hat T_{\!\sigma}^{\,\sigma}g^{\mu\nu}\big)\, , \eqn{2}\fe
where $\hat T^{\mu\nu}$ is the stress momentum energy density 
tensor of the source.

The problem of ultraviolet divergences for point particle or string models
arises because in these cases the relevant source densities are not regular
functions but Dirac type distributions that vanish outside the relevant
one or two dimensional world sheets. In the case of a string with local
worldsheet imbedding given by $x^\mu= \bar x^\mu\{\sigma\}$ in terms of
intrinsic coordinates $\sigma^i$ ($i=0,1$), so that the induced surface
metric will have the form $  \gamma_{ij}= g_{\mu\nu}\bar x{^\mu}{_{\! ,i}}\,
\bar x{^\nu}{_{\!,j}}$, the relevant source distribution will be expressible
using the terminology of Dirac delta ``functions'' in the form
\be \hat T{^{\mu\nu}}=\Vert g\Vert^{-1/2}\int \overline T{^{\mu\nu}}\, 
\delta^{\rm 4}[x-\bar x\{\sigma\}]\, \Vert\gamma \Vert^{1/2}
\, d^2\sigma\, ,\eqn{4}\fe
where $\vert\gamma\vert$ is the determinant of the induced metric,
and the surface stress energy density $\overline T{^{\mu\nu}}$,  is a 
{\it regular} tensorial function on the worldsheet (but undefined off it).

In the simple elastic models considered here, the only internal field on 
the string will be a surface current density 
$ \overline c{^\mu}=\varepsilon^{ij} \bar x{^\mu}_{,i} \psi_{, j}$,
where  $\psi$ is a scalar stream function on the world sheet (which will be a
free variable in the variation formulation described below), using the
notation $\varepsilon^{ij}=-\varepsilon^{ji}$ for the antisymmetric worldsheet
measure tensor that is specified (modulo a choice of sign representing an
orientation convention) as the square root of the induced metric, i.e.
$\varepsilon^{ik}\varepsilon_{kj}=\gamma^i_{\ j}$. If there were a non-zero
charge coupling constant $q$, as supposed in  Witten's theory~\cite{W85} of
superconducting strings, then this would correspond to an electric surface
current density $\overline j{^\mu}=q\,\overline c{^\mu}$. However the
present discussion is concerned just with the early regime in which effects
of electromagnetic coupling are negligible compared with those of
gravitation. This means that in the relevant action for the ``bare'' string
model (i.e. the non self interacting limit), as given by  an integral of the
form,
${\cal I}=\int \overline{\cal L}\, \Vert\gamma\Vert^{1/2}\, d^2\sigma$,
the specification of the relevant Lagrangian scalar $\overline{\cal
L}$ on the worldsheet will be given by a master function, $\Lambda$, that 
depends only on the undifferentiated background metric and the gradient
of the stream function $\psi$, according the formula
\be\overline{\cal L}=\Lambda 
+{_1\over^2} \overline T{^{\mu\nu}} h_{\mu\nu}
\, ,\eqn{60}\fe
in which the coefficient for the linearised gravitational adjustment term
here is the surface stress-energy tensor that specifies the gravitational
source in (\ref{2}), which is given by
$\overline T{^{\mu\nu}}= 2\Vert\gamma\Vert^{-1/2}
{\partial \big(\Lambda\Vert\gamma\Vert^{1/2}\big)/ \partial g_{\mu\nu}}$.

Since the ensuing field equations will evidently involve gradients of the
stress energy tensor, their formulation will require the introduction of the
appropriately defined hyper-Cauchy tensor (a relativistic generalisation
of the Cauchy elasticity tensor of classical mechanics) which is defined by
$\overline{\cal C}{^{\mu\nu\rho\sigma}}=$ $\Vert\gamma\Vert^{-1/2}
\partial \big(\overline T{^{\rho\sigma}}
\Vert\gamma\Vert^{1/2}\big)/\partial g_{\mu\nu}$.
In terms of this quantity, the dynamical equations obtained from the
Lagrangian (\ref{60}) can be shown~\cite{BC95} to be expressible in the
standard form 
\be \overline\nabla_{\!\nu}\overline T{^{\mu\nu}}
= f_{\rm g}{^\mu}\, ,\eqn{20}\fe 
where the effective gravitational force density vector is given by
\be f_{\rm g}{^\mu}= {_1\over^2}\overline T{^{\nu\sigma}}
\nabla^\mu h_{\nu\sigma} - \overline\nabla_{\!\nu}\big(
\overline T{^{\nu\sigma}} h_\sigma{^\mu}+
\overline{\cal C}{^{\mu\nu\rho\sigma}} h_{\rho\sigma} \big)
\, , \eqn{22}\fe
using the notation $ \overline\nabla_{\!\mu} =
\eta_\mu^{\ \nu}\nabla_{\!\nu}\, $
for the tangentially projected gradient operator, where  $\eta_\mu^{\ \nu}$
is the tangential projection tensor, i.e. the index lowered form of the
first fundamental tensor of the worldsheet, which is obtained simply by
mapping its internal metric onto the spacetime background according to the
formula
$ \eta^{\mu\nu}=\gamma^{ij}\bar x{^\mu}{_{\! ,i}}\, 
\bar x{^\nu}{_{\! ,j}}\, .$ 

\bigskip
\section{Renormalisation}

The problem with the application of (\ref{22}) is of course that the linearised
gravitational field will consist not just of a well behaved long range
contribution, $\widetilde h_{\mu\nu}$ say, but also of a divergent short range
self interaction contribution, $\widehat h_{\mu\nu}=$ $ h_{\mu\nu}-\widetilde
h_{\mu\nu}$, that needs to be appropriately regularised in the manner recently
described in the analogous electromagnetic case~\cite{C97}. This routine
procedure leads to a result that is proportional to the relevant source in
(\ref{2}), which gives
\be\widehat h_{\mu\nu}=2\hbox{\sm G}\ \widehat l\
\big(2\overline T_{\!\mu\nu}-
\overline T_{\!\sigma}{^\sigma}g_{\mu\nu}\big)\, ,\eqn{33}\fe
where, as usual for a string self interaction in 4 dimensions, the
proportionality factor has the form 
\be \widehat l= {\rm ln}\big\{ {\Delta^2
/\delta_\ast^{\, 2} } \big\} \eqn{33a}\fe 
in terms of an ``ultraviolet'' cut off
lengthscale $\delta_\ast$  representing the effective thickness of the string,
and a much larger ``infrared'' cut off $\Delta$, given by a lengthscale
characterising the large scale geometry of the string configuration. As
pointed out in the electromagnetic case~\cite{C97}, the corresponding
regularised value of the gradient of such a divergent self field will be
obtainable from the regularised self field by the application of the
regularised gradient operator defined by 
\be \widehat\nabla_{\!\mu}=\overline\nabla_{\!\mu}+{_1\over^2}K_\mu
\, ,\eqn{34}\fe 
where $K_\mu$ is the worldsheet curvature vector that is obtainable as the
surface divergence of the fundamental (tangential projection) tensor,
$K_\mu=\overline\nabla_{\!\nu}\eta^\nu_{\, \mu}$.

When the corresponding divergent self force contribution is evaluated
substituting (\ref{33}) and (\ref{34}) in (\ref{22}), the result turns out,
rather remarkably, to be describable as a renormalisation of the stress
energy tensor, since one obtains~\cite{C98} a regularised self force that is
expressible as world sheet divergence 
\be\widehat f_{\rm g}^{\,\mu}=-\overline\nabla_{\!\nu}
\widehat T_{\!\rm g}{^{\mu\nu}}
\, ,\eqn{39}\fe
in which the relevant stress momentum energy density contribution from the
gravitational self interaction has the form
\be\widehat T_{\!\rm g}{^{\mu\nu}}=\widehat h_\sigma{^\mu}
\overline T{^{\nu\sigma}} 
-{_1\over^4}\widehat h_{\rho\sigma}\overline T{^{\rho\sigma}}\eta^{\mu\nu}
+\widehat h_{\rho\sigma}\overline{\cal C}{^{\rho\sigma\mu\nu}}
\, .\eqn{40}\fe
It is also to be observed that this ``on shell'' self gravitational stress
energy contribution is obtainable from a corresponding ``off shell'' self
gravitational action contribution given by
\be \widehat\Lambda_{\rm g}= {_1\over ^4}\overline T{^{\mu\nu}}
\widehat h_{\mu\nu}={_1\over^2}\hbox{\sm G}\, \widehat l\,
\big( 2\overline T_{\!\mu\nu}\overline T{^{\mu\nu}}
-\overline T_{\!\mu}{^\mu}\overline T_{\!\nu}{^\nu}\big)
\, .\eqn{67a}\fe
This provides a regularised treatment in which the original worldsheet
Lagrangian ${\cal L}$  is replaced by a regularised Lagrangian
$\widetilde{\cal L}=\widetilde\Lambda +{1\over 2} \overline
T{^{\mu\nu}\widetilde h_{\mu\nu}}$ involving only the well behaved part
$\widetilde h_{\mu\nu}$ of the gravitational field, whereby the divergent part
$\widehat h_{\mu\nu}$ is absorbed into a renormalised master function given by
\be
\widetilde\Lambda=\Lambda + \widehat\Lambda_{\rm g}\, .\eqn{70}\fe
The consistency of this treatment has been neatly confirmed by an
independent investigation in which, by working entirely at the level of the
``off shell'' action in a space-time of arbitrary dimension, Buonanno and
Damour~\cite{BD98} have recently obtained a general self-interaction formula
whose specialisation to the case of gravitation in 4-dimensions is in
precise agreement with the result (\ref{67a}) obtained here.

For any simple elastic string model of the kind considered here, the master
function $\Lambda$ will depend just on the scalar magnitude that is
specifiable~\cite{CP95} as $\chi=\overline c^\mu \overline c_\mu=$
$-\gamma^{ij}\psi_{, i}\psi_{,j}= -p_\mu p^\mu$ where the relevant momentum
vector is defined by
$p^\mu=\overline\nabla{^\mu}\psi=\bar x{^\mu}_{\, , i}\gamma^{ij}\psi_{,j}$.
Using the notation $\Lambda^\prime=d\Lambda/d\chi$, 
it can be seen that the regularised self field will be given by
\be \widehat h^{\mu\nu}= 4\hbox{\sm G}\, \widehat l\,\big(
2\Lambda^\prime p^\mu p^\nu+\chi\Lambda^\prime g^{\mu\nu}-
\Lambda\perp^{\!\mu\nu}\big)\, ,\fe
using the notation $\perp^{\!\mu}_{\,\nu}=g^\mu_{\ \nu}-\eta^\mu_{\ \nu}$
for the (rank 2) worldsheet orthogonal projection tensor. The corresponding
self gravitational action contribution (\ref{67a}) will be expressible as
\be \widehat\Lambda_{\rm g}={_1\over^2}\hbox{\sm G}\, \widehat l\,
\big(U-T\big)^2=
         2\hbox{\sm G}\ \widehat l\,\big(\chi\Lambda^\prime\big)^2
\eqn{67b}\, .\fe

\bigskip
\section{Effect on stress tensor and propagation speeds}

In analogy with the standard formula
\be \overline T{^{\mu\nu}}=\Lambda\eta^{\mu\nu}+2\Lambda^\prime p^\mu
p^\nu \, ,\fe
for the ``bare'' surface stress energy tensor, the corresponding self
gravitational stress energy tensor (\ref{40}) will be obtainable in the form
\be \widehat T_{\!\rm g}{^{\mu\nu}}=
\widehat\Lambda_{\rm g}\eta^{\mu\nu}+2
\widehat\Lambda_{\rm g}^{\,\prime}\, p^\mu
p^\nu \, .\fe

If the current is timelike -- as can be assumed without loss of generality
in the wiggly Nambu-Goto string approximation -- so that we have $\chi\leq
0$, then the energy density and tension will be given for the bare model by
$U=-\Lambda$ and $T= 2\chi\Lambda^\prime-\Lambda$,
while the corresponding extrinsic (wiggle type) and longitudinal
(sound type) perturbations speeds $c_{_{\rm E}}$ and $c_{_{\rm L}}$
will be given~\cite{CP95} by
\be c_{_{\rm E}}^{\, 2}=1-2\chi\Lambda^\prime/\Lambda\, ,\hskip 1 cm
c_{_{\rm L}}^{\, 2}= 
1+2\chi\Lambda^{\prime\prime}/\Lambda^\prime\, .\eqn{80}\fe
An elastic string state is describable as supersonic, transonic, or subsonic,
according to whether the difference,
\be c_{_{\rm E}}^{\, 2}-c_{_{\rm L}}^{\, 2}=
- 2\chi\,\Big(\ln\{\Lambda\Lambda^\prime\}\Big)^\prime\, ,\eqn{81}\fe
is positive, zero, or negative.
In terms of these quantities the corresponding renormalised energy density
(for a given value of the current magnitude as specified by $\chi$) will be
obtainable directly from the renormalised action (\ref{70}) as
\be \widetilde U=-\widetilde \Lambda =U -
{_1\over^2}\hbox{\sm G}\, \widehat l\ UT\,
\big(c_{_{\rm E}}^{\,-1}-c_{_{\rm E}}\big)^2\, .\fe
However the evaluation of the corresponding renormalised tension is not so 
quite so simple: it works out to be
\be \widetilde T=T+{_1\over^2}\hbox{\sm G}\, \widehat l\ UT\,
\big(c_{_{\rm E}}^{\,-1}-c_{_{\rm E}}\big)^2\big(1+2c_{_{\rm L}}^{\, 2}\big) \,
 .\fe

\bigskip
\section{The case of the transonic wiggly string model} 

In the exact Nambu-Goto case~\cite{CB98} the master function is just a
constant, $\Lambda=-m^2$, where 
(in units with the Dirac Planck constant $\hbar$ set to unity)
the parameter $m$ is the relevant Kibble mass scale,
which is a constant that can be expected to be of the same order of
magnitude as the Higgs mass scale associated with the underlying
symmetry breaking for a string of the ordinary ``cosmic'' kind,
representing a vortex type defect of the vacuum. In this case one
simply obtains $\Lambda^\prime=0$, so the divergent short range self
gravitational contribution vanishes.

However for the purpose of a course grained description on larger
scale, it is appropriate to use the transonic model~\cite{C90,V90} to
represent the smoothed average over shortwavelength wiggles in the
underlying microscopic Nambu-Goto model.  For this transonic string
model, the relevant master function has the form 
\be
\Lambda=-m\sqrt{m^2-\chi}\, ,\eqn{82}\fe 
with non zero derivative
\be \Lambda^\prime=-m^2/2\Lambda\, .\eqn{83}\fe 
This gives a non-zero self gravitional contribution attributable to 
interaction at intermediate range, i.e. distances large compared with the 
wiggle wavelength in the underlying macroscopic model but small compared 
with the smoothing length on which the macroscopic description is based. 
A noteworthy example of the application of this model is to the 
case of wiggles that are of purely thermal origin, for which
the state parameter $\chi$ will be given\cite{C90,C94} as a function of 
the relevant temperature $\Theta$ by the formula
\be
\chi={-2\pi m^2\Theta^2\over 3m^2- 2\pi\Theta^2}\, ,\eqn{84}\fe
from which it can be seen that corresponding wiggle propagation speed
will be given by
\be
c_{_{\rm E}}^{\, 2}=1-{2\pi\Theta^2\over 3m^2} \, .\eqn{85}\fe

It is evident that the special transonicity property 
\be
c_{_{\rm L}}=c_{_{\rm E}} \eqn{86}\fe
of the ``bare''  model will not survive in the renormalised model, as obtained
using (\ref{82}) from (\ref{70}) and (\ref{67b}), for which (still working
just to linear order in the gravitational coupling) the difference between the
squared propagation speeds is found to be given by
\be \widetilde{ c_{_{\rm E}}^{\, 2}} -\widetilde{ c_{_{\rm L}}^{\, 2}} =
\hbox{\sm G}\, m^2\, \widehat l\ (c_{_{\rm E}}^{-1}
-c_{_{\rm E}})\Big((1+c_{_{\rm E}}^{\, 2})^2
+{_1\over ^2}(1-c_{_{\rm E}}^{\, 2})^2\Big) \, .\eqn{87} \fe
The manifest positivity (since $0<c_{_{\rm E}}< 1$) of this result shows
that the ``dressed'' gravitationally self interacting wiggly string model 
will be of supersonic type.

\section{Orders of magnitude.}

The regime of applicability of the foregoing analysis is
of course subject to limitations. To start with, for the
validity of the linearised gravitation equation (\ref{2}) on
which the entire analysis depends, the weak coupling condition
\be 
\hbox{\sm G}\, m^2\, \widehat l\, \ll 1 \eqn{50}\fe
must be satisfied. This requirement is not very restrictive,
because even the heaviest kind cosmic strings that are commonly
considered in cosmological applications, namely those arising
from G.U.T. symmetry breaking, are characterised by
$\hbox{\sm G} m^2\approx 10^{-6}$. For other kinds, such as those
arising from electroweak symmetry breaking, the value
of $\hbox{\sm G} m^2$ will be  smaller still. The smallness of
$\hbox{\sm G} m^2$ will be partially counterbalanced by the fact 
that the regularisation factor $\widehat l$ will be large compared with 
unity, but since, according to (\ref{33a}), it arises as a logarithm, it 
can never be extremely large. In nearly all cases that are likely
to arise in practice it will satisfy $\widehat l\ll 10^2$, so the
requirement (\ref{50}) will be satisfied by a large margin.

Since the rigourous justification~\cite{C95}
of the description of the macroscopic effect of the wiggles
by the simple elastic model (\ref{82}) depends on the supposition that self
intersections of the string can be neglected (since if loop formation
were important a more elaborate non-elastic model would be needed)
the validity of the model as a precise representation is limited
to the regime in which the effective energy density of the wiggles,
as formally defined by
\be
\varepsilon={U-T\over 2}\simeq U-m^2 \, ,\eqn{88}\fe
is small compared with the intrinsic energy density of the string,
i.e.
\be 
\varepsilon\ll m^2\, ,\eqn{89} \fe
which is interpretable in the thermal case as meaning
\be \Theta \ll m\, .\fe
Since this is equivalent to the restriction
\be 1 - c_{_{\rm E}}^{\, 2}\ll 1 \eqn{90} \fe
it can be seen that
the difference (\ref{87}) will be given approximately by
the simple formula 
\be 
\widetilde{ c_{_{\rm E}}^{\, 2}} -\widetilde{ c_{_{\rm L}}^{\, 2}} 
\simeq 8\, \hbox{\sm G}\ \widehat l\, \varepsilon
\ .\eqn{91}\fe

In an application of this kind, the magnitude, $\lambda$ say, typifying
the wavelength of the wiggles over which the averaging is taken will
provide an appropriate choice for the infra-red cut off, i.e. it will
be natural to take $\Delta\approx\lambda$. Similarly the magnitude,
$\alpha$ say, characterising the amplitude of the wiggles will provide
the corresponding value for the ultra-violet cut off, which will thus
be given by $\delta_\ast\approx\alpha$, provided that,
as will usually be the case, the amplitude of the wiggles is not even
smaller than the microscopic string radius, $r$ say, in which case the
latter would itself provide the relevant ultraviolet cut off.  When the
string is of the usual ``cosmic'' variety, representing an underlying
vortex type defect of the vacuum, one expects the radius to be of the
same order of magnitude as the Compton wavelength associated with the
relevant Kibble mass, i.e. one expects to have $r\approx m^{-1}$.  
In most applications of interest, the relevant
amplitude $\alpha$ will be considerably larger than this, and even for
wiggles of purely thermal origin it will never be smaller:  for the
wiggles produced by a given temperature $\Theta$, as given by (\ref{84}),
one can estimate the relevant magnitudes as
$\lambda\approx\Theta^{-1}$ and $\alpha\approx m^{-1}$, i.e.
independently of the temperature the relevant amplitude will be of the
same order as the cosmic string radius, $\alpha\approx r$. This means
that, in the usual physical applications, it will generally be possible
to take the relevant cut-off ratio to be given by
$\Delta/\delta_\ast\approx\lambda/\alpha$.

Since the effective energy density, $\varepsilon$, contributed
by small wiggles of wavelength $\lambda$ and amplitude $\alpha$ can 
be roughly estimated as $\varepsilon\approx m^2\alpha^2/\lambda^2$, we 
see that the squared cut-off ratio appearing in the logarithm in 
(\ref{33a}) will be given  by the corresponding rough estimate
$\Delta^2/\delta_\ast^{\, 2} \approx m^2/\varepsilon$,
and hence that a reasonably accurate description should be
obtainable by taking  the regularisation factor itself to be given
by an approximation of the form
\be 
\widehat l\simeq {\rm ln}\Big\{ {m^2\over \langle\varepsilon\rangle} 
\Big\} \, ,\eqn{93}\fe 
where $\langle\varepsilon\rangle$ is a constant chosen
as some suitably weighted mean value of the energy density
$\varepsilon=m^2\chi/2\Lambda$ in the string segment under consideration.
The requirement (\ref{89}) that the wiggle energy density should be 
small compared with the intrinsic energy of the string implies that
the ensuing factor $\widehat l$ will be reasonably large compared with unity,
and the consideration that the dependence is logarithmic means that
the result will be inensitive to the details of the particular prescription
chosen to specify $\langle\varepsilon\rangle$. In order to obtain higher
accuracy , one might be tempted to replace the fixed mean value
$\langle\varepsilon\rangle$ in (\ref{93}) by the variable local value
of $\varepsilon$,  which, by (\ref{67b}), would be equivalent to taking
the self gravitational action adjustment to be
\be 
\widehat\Lambda_{\rm g} \simeq 2 \hbox{\sm G}\,
\varepsilon^2\,  {\rm ln}\Big\{ {m^2\over \varepsilon} \Big\}
 \, \eqn{94}\fe
with
\be
\varepsilon={-m\chi\over2\sqrt{m^2-\chi}}
\, .\fe
However the appearance of improvement provided by such use of a
variable rather than constant value for the renormalisation factor
$\widehat l$ is rather illusory, since the preceeding demonstration of
renormalisability -- as embodied in the formulae (39), (40) and (67a)
-- was dependent on the postulate that  $\widehat l$ should be
constant. Moreover it can easily be checked explicitly that the only
effect of the use of the apparently more precise formula (\ref{94}) on
the final formula (\ref{91}) will be to replace the factor $\widehat l$
by $\widehat l-{3\over 2}$. This adjustment would be significant only
if $\epsilon$ were so large as to be comparable with $m^2$, so that the
accuracy of the treatment would in any case be affected by other
complicating processess such as loop formation due to self
intersections.

\vfill\eject

The author wishes to thank Xavier Martin, Paul Shellard, and Alex Vilenkin 
for earlier discussions about the treatment of string wiggles and to thank 
Richard Battye, Thibault Damour, Alejandro Gangui, Gary Gibbons, and 
Patrick Peter for more recent discussions about regularisation and 
renormalisation.


\begin{thebibliography}{99}

\bibitem{VS94} A. Vilenkin, E.P.S. Shellard (1994)
{\it Cosmic strings and other topological defects}
(Cambridge University Press).

\bibitem{W85} E. Witten (1985)
{\it Nucl. Phys.} {\bf B249}, 557.

\bibitem{CP95} B. Carter, P. Peter (1995)
{\it Phys. Rev} {\bf D52}, R1744.
[hep-ph/9411425]

\bibitem{CM93} B. Carter, X. Martin (1993)
{\it Ann. Phys.} {\bf 227}, 151. 

\bibitem{CPG97} B. Carter, P. Peter, A. Gangui (1997)
{\it Phys. Rev.} {\bf D55}, 4647.
[hep-ph/9609 401].

\bibitem{DS88} R.L. Davis, E.P.S Shellard (1988)
{\it Phys. Letters} {bf B 209}, 485.

\bibitem{SA90} E.P.S. Shellard, B. Allen (1990)
in {\it The formation and evolution of cosmic strings}, ed. G. Gibbons, 
S.W. Hawking, T. Vachaspati (Cambridge University Press) {\it pp} 421-448.

\bibitem{C90} B. Carter (1990)
{\it Phys. Rev.} {\bf D 41}, 3869. 

\bibitem{V90} A. Vilenkin (1990)
{\it Phys. Rev.} {\bf D41}, 3038.

\bibitem {M95} X. Martin (1995)
{\it Phys. Rev. Lett.} {\bf 74}, 3102.

\bibitem{C95} B. Carter (1995)
{\it Phys. Rev. Lett.} {\bf 74}, 3098. [hep-th/9411231]

\bibitem{C97} B. Carter (1997) 
{\it Phys. Letters} {\bf B404}, 246. [hep-th/9704210] 


\bibitem{GPB98} A. Gangui, P. Peter, C. Boehm (1998)
{\it Phys. Rev.} {\bf D57}, 2580. [hep-ph/9705204]

\bibitem{BC95} R.A. Battye, B. Carter (1995)
{\it Phys. Letters} {\bf B357}, 29. [hep-th/9507059]

\bibitem{C98} B. Carter (1999)
in {\it Strings, branes, and dualities} (NATO ASI {\bf C520}, Carg\`ese, 1997),
ed L. Baulieu, P. Di Francesco, M. Douglas, V. Kazakov, M. Pico,
P. Windey (Kluwer, Dordrecht) {\it pp} 441-444.
[hep-th/9802019]

\bibitem{CB98} B. Carter, R.A. Battye (1998)
{\it Phys. Letters} {\bf B430}, 49-53. [hep-th/9803012]

\bibitem{BD98} A. Buonanno, T.Damour (1998)
{\it Phys. Letters} {\bf B432}, 51-57. [hep-th/9803025]

\bibitem{G98} G. W. Gibbons (1998)
``Branes as BIons''
[hep-th/9803203]

\bibitem{C94} B. Carter (1994)
 {\it Nuclear Physics} {\bf B412}, 345-371.

\end{thebibliography}
\end{document}